\documentclass[a4paper,reprint,12pt,aps,tightenlines,onecolumn,notitlepage]{revtex4-1}
\usepackage{amsmath}
\usepackage{amssymb}
\usepackage{cancel}
\usepackage{stmaryrd}
\usepackage{amsfonts}
\usepackage[T1]{fontenc}
\usepackage{bm}
\usepackage{color}
\usepackage{graphicx}
\DeclareSymbolFont{operators}{OT1}{cmr}{m}{n}
\DeclareSymbolFont{letters}{OML}{cmm}{m}{it}
\DeclareSymbolFont{symbols}{OMS}{cmsy}{m}{n}
\DeclareSymbolFont{largesymbols}{OMX}{cmex}{m}{n}
\usepackage[hyperindex,breaklinks]{hyperref}
\hypersetup{
    pdfborder={0 0 0},
    allbordercolors={0 0 0},
    bookmarks=true,         
    unicode=false,          
    pdftoolbar=true,        
    pdfmenubar=true,        
    pdffitwindow=false,     
    pdfstartview={FitH},    
    pdfcreator={Eero Hirvijoki},
    pdfproducer={pdflatex},
    pdfnewwindow=true,
    colorlinks=true,
    linkcolor=blue,
    citecolor=blue,
    filecolor=blue,
    urlcolor=blue,  
}

\DeclareFontFamily{OT1}{pzc}{}
\DeclareFontShape{OT1}{pzc}{m}{it}{<-> s * [1.10] pzcmi7t}{}
\DeclareMathAlphabet{\mathpzc}{OT1}{pzc}{m}{it}

\makeatother

\begin{document}
\title{Conservative finite-element method for the relativistic Coulomb collision operator}
\author{Eero Hirvijoki}
\affiliation{Aalto University, Department of Applied Physics, P.O.Box 11100, FI-00076, Finland}
\email{eero.hirvijoki@gmail.com}

\date{\today}

\begin{abstract}
    This research note documents new developments regarding finite-element discretizations of the relativistic Beliaev-Budker Coulomb collision operator and the nonrelativistic Landau operator. Where energy conservation in a finite-element approximation of the relativistic collision operator was previously thought to be elusive, it is now achieved even with linear elements. The same result applies to the nonrelativistic Landau operator for which the energy conservation was thought to require at least quadratic elements. In both cases, the momentum and density conservation are guaranteed as previously. The new outcomes benefit from the findings reported in a recent finite-difference-scheme paper [Shiroto \& Sentoku, arXiv:1902.07866] which we generalize to the finite-element method. This note focuses solely on the direct discretization of the collision operator, leaving the discretization of the underlying metriplectic formulation of the relativistic collision operator to future publications.
\end{abstract}

\maketitle

{\it Introduction: }
Structure-preserving numerical schemes have become a topic of intense discussion and development in the recent years. On this front, the Coulomb collision operator, based on the Landau approximation~\cite{Landau:1936}, has received attention from finite-element, finite-difference, and mixed schemes, with conservation properties varying from the invariant moments to positivity-preserving and entropic schemes. Both Landau's original version and the so-called Rosenbluth potential formulation~\cite{RMJ1957} have been discussed in detail. For further discussion, we encourage the reader to consult, e.g., the papers \cite{Lemou2005ImplicitSF,Buet_LeThanh:hal-00092543,Yoon:2014:POP,Taitano:2015JCP,Hager:2016:JCP,Hirvijoki-Adams:2017PhPl,Kraus-Hirvijoki-2017,Hirvijoki-Kraus-Burby:2018arXiv,Hirvijoki-Burby-Kraus:2018arXiv}.

Nevertheless, structure-preserving discretization of the relativistic Beliaev-Budker operator~\cite{Beliaev_Budker_1956} has turned out to be somewhat elusive. While non-conservative solvers, relying on the potential formulation~\cite{Braams-Karney:PRL1987}, have been implemented and used in production level simulations~\cite{Stahl-et-al:2017CoPhC}, it took until late 2018 for the first conservative discretization, relying on the potential formulation and enforced nonlinear constraints, to be reported in the APS-DPP meeting~\cite{Daniel-Taitano-Chacon:arXiv:1902.10241}. Only very recently have the symmetries of the original, integral formulation of the collision operator been realized in a finite-difference scheme~\cite{shiroto-sentoku:2019arXiv}. In this note we focus on these intrinsic symmetries and generalize the idea reported in the finite-difference paper~\cite{shiroto-sentoku:2019arXiv} to finite-element methods. While at work, we discuss also how the energy conservation of the nonrelativistic Landau operator can be achieved with only linear elements. Previously, this was thought to be possible only with quadratic or higher-order elements~\cite{Hirvijoki-Adams:2017PhPl}.

{\it The collision operator: }
Both the nonrelativistic Landau operator and the relativistic Beliaev-Budker operator for species $a$ colliding with species $b$ can be written in the form
\begin{align}
\frac{\partial f_a}{\partial t}=\frac{c_{ab}}{m_a}\frac{\partial}{\partial\bm{u}}\cdot\int \mathbf{Q}(\bm{u},\bm{u}')\cdot\bm{\Gamma}_{ab}(\bm{u},\bm{u}')d\bm{u}',
\end{align}
where the symmetric scalar coefficient $c_{ab}$ is given by
\begin{align}
c_{ab}=\frac{e_a^2e_b^2}{8\pi\varepsilon_0^2}\ln\Lambda_{ab},
\end{align}
with $e_a$ ($e_b$) being the species $a$ ($b$) charge, and the anti-symmetric vector $\bm{\Gamma}_{ab}(\bm{u},\bm{u}')$ is
\begin{align}
\bm{\Gamma}_{ab}(\bm{u},\bm{u}')=\frac{f_b(\bm{u}')}{m_a}\frac{\partial f_a}{\partial\bm{u}}-\frac{f_a(\bm{u})}{m_b}\frac{\partial f_b}{\partial\bm{u}'}=-\bm{\Gamma}_{ba}(\bm{u}',\bm{u}).
\end{align}
Note that the species $b$ could also be the same as species $a$ and that the operator could be a sum over multiple different $b$ species.

The possibility of dealing with relativistic phenomena relates to choosing the tensor $\mathbf{Q}(\bm{u},\bm{u}')$ correctly. The version derived by Beliaev and Budker, namely
\begin{align}
\mathbf{Q}_{\text{BB}}(\bm{u},\bm{u}')=\frac{r^2}{\gamma\gamma'w^3}\left(w^2\mathbf{1}-\bm{u}\bm{u}-\bm{u}'\bm{u}'+r(\bm{u}\bm{u}'+\bm{u}'\bm{u})\right),
\end{align}
with $\gamma(\bm{u})=\sqrt{1+\bm{u}^2/c^2}$, $r=\gamma\gamma'-\bm{u}\cdot\bm{u}'/c^2$, $w=c\sqrt{r^2-1}$, and $\bm{u}=\bm{p}/m_a=\gamma\bm{v}$ and $\bm{u}'=\bm{p}'/m_b=\gamma'\bm{v}'$, accounts for relativistic velocities of the colliding particles whereas Landau's nonrelativistic version, with
\begin{align}
    \mathbf{Q}_{\text{L}}(\bm{v},\bm{v}')=\frac{1}{|\bm{v}-\bm{v}'|}\left(\mathbf{1}-\frac{(\bm{v}-\bm{v}')(\bm{v}-\bm{v}')}{|\bm{v}-\bm{v}'|^2}\right),
\end{align}
is the limit of the Beliaev-Budker tensor at $c\rightarrow\infty$ with $\bm{u}\rightarrow\bm{v}$ and $\bm{u}'\rightarrow\bm{v}'$. Although $\bm{u}$ and $\bm{v}$ have different meaning in the relativistic context, we will use $\bm{u}$, interpreted as $\bm{v}$, in case of the nonrelativistic Landau operator, simply to unify the discussion and to avoid extra clutter.

The conservation properties of both operators follow in a similar fashion: Multiply the collision operator of species $a$ with a test function $\psi_a(\bm{u})$ and integrate the expression over the space $\bm{u}$, leading to the weak expression
\begin{align}\label{eq:weak-form}
\int \psi_a \frac{\partial f_a}{\partial t} d\bm{u}=-c_{ab}\int\int \frac{1}{m_a}\frac{\partial\psi_a}{\partial\bm{u}}\cdot \mathbf{Q}(\bm{u},\bm{u}')\cdot\bm{\Gamma}_{ab}(\bm{u},\bm{u}')d\bm{u}'d\bm{u}.
\end{align}
Write similarly for species $b$ colliding with species $a$ to obtain
\begin{align}
\int \psi_b \frac{\partial f_b}{\partial t} d\bm{u}'
=-c_{ba}\int\int \frac{1}{m_b}\frac{\partial\psi_b}{\partial\bm{u}'}\cdot \mathbf{Q}(\bm{u}',\bm{u})\cdot\bm{\Gamma}_{ba}(\bm{u}',\bm{u})d\bm{u}d\bm{u}'.
\end{align}
Then, use the symmetry of $\mathbf{Q}(\bm{u}',\bm{u})=\mathbf{Q}(\bm{u},\bm{u}')$ and $c_{ba}=c_{ab}$, and the antisymmetry of  $\bm{\Gamma}_{ba}(\bm{u}',\bm{u})=-\bm{\Gamma}_{ab}(\bm{u},\bm{u}')$ to obtain
\begin{align}\label{eq:conservation_properties}
&\int \psi_a \frac{\partial f_a}{\partial t} d\bm{u}+\int \psi_b \frac{\partial f_b}{\partial t} d\bm{u}'\nonumber\\
&=-c_{ab}\int\int \left(\frac{1}{m_a}\frac{\partial\psi_a}{\partial\bm{u}}-\frac{1}{m_b}\frac{\partial\psi_b}{\partial\bm{u}'}\right)\cdot \mathbf{Q}(\bm{u},\bm{u}')\cdot\bm{\Gamma}_{ab}(\bm{u},\bm{u}')d\bm{u}'d\bm{u}.
\end{align}
In the relativistic case, one chooses $\psi_a=m_a\{1,u^x,u^y,u^z,\gamma c^2\}$ and $\mathbf{Q}=\mathbf{Q}_{\text{BB}}$ to see that the number, momentum, and energy density are conserved. In the nonrelativistic case the corresponding choices are $\psi_a=m_a\{1,u^x,u^y,u^z,|\bm{u}|^2/2\}$ and $\mathbf{Q}=\mathbf{Q}_{\text{L}}$. In both cases the number and momentum density conservation follow from the expression
$$
\frac{1}{m_a}\frac{\partial\psi_a}{\partial\bm{u}}-\frac{1}{m_b}\frac{\partial\psi_b}{\partial\bm{u}'}
$$
vanishing identically while the energy conservation exploits the null space of the tensor $\mathbf{Q}$. In the relativistic case one has
\begin{align}\label{eq:relativistic_null_space}
\left(\frac{\partial\gamma c^2}{\partial\bm{u}}-\frac{\partial\gamma'c^2}{\partial\bm{u}'}\right)\cdot\mathbf{Q}_{\text{BB}}(\bm{u},\bm{u}')=\left(\frac{\bm{u}}{\gamma}-\frac{\bm{u}'}{\gamma'}\right)\cdot\mathbf{Q}_{\text{BB}}(\bm{u},\bm{u}')=\bm{0},
\end{align}
and in the nonrelativistic case the corresponding result is
\begin{align}\label{eq:nonrelativistic_null_space}
\left(\frac{\partial \tfrac{1}{2}|\bm{u}|^2}{\partial\bm{u}}-\frac{\partial \tfrac{1}{2}|\bm{u}'|^2}{\partial\bm{u}'}\right)\cdot\mathbf{Q}_{\text{L}}(\bm{u},\bm{u}')=(\bm{u}-\bm{u}')\cdot\mathbf{Q}_{\text{L}}(\bm{u},\bm{u}')=\bm{0}.
\end{align}

{\it Finite-element approach: } Next we choose two sets of basis functions $\{\phi^a_i\}_{i\in I^a}$ and $\{\phi^b_i\}_{i\in I^b}$, one set for species $a$ and the other for species $b$. The function sets could be the same for both species, but since the masses of electrons and ions are very different it makes sense to have the possibility for different phase-space domains for different species. Similarly, we assume that there are quadrature points and weights according to $\{\bm{\xi}^a_{p},w^a_p\}_{p\in P^a}$ and $\{\bm{\xi}^b_{p},w^b_p\}_{p\in P^b}$ for performing integrals numerically over the domains the bases cover. The finite-element discretizations of the collision operators of species $a$ and $b$ are then obtained from the corresponding weak expressions by choosing the test functions $\psi_a$ and $\psi_b$ from the sets $\{\phi^a_{i}\}_{i\in I^a}$ and $\{\phi^b_{i}\}_{i\in I^b}$, substituting $f_a(t,\bm{u})=f_a^i(t)\phi^a_i(\bm{u})$ and $f_b(t,\bm{u}')=f_b^j(t)\phi^b_j(\bm{u}')$, and performing the integrals numerically with the given quadrature points and weights. This leads to ordinary differential equations for the degrees of freedom $\{f_a^i\}_{i\in I^a}$ and $\{f_b^i\}_{i\in I^b}$, given by
\begin{align}\label{eq:equations_of_motion_fa}
    \sum_{p,i}w_p^a\phi^a_k(\bm{\xi}^a_{p})\phi^a_i(\bm{\xi}^a_{p})\frac{\partial f_a^i}{\partial t}&=-\sum_{p,q}w_p^aw_q^b\frac{c_{ab}}{m_a}\frac{\partial\phi^a_k}{\partial\bm{u}}\Big\vert_{\bm{\xi}^a_p}\cdot \mathbf{Q}(\bm{\xi}^a_p,\bm{\xi}^b_q)\cdot\bm{\Gamma}_{ab}(\bm{\xi}^a_p,\bm{\xi}^b_q),\quad \forall k\in I^a,\\
    \label{eq:equations_of_motion_fb}
    \sum_{q,j}w_q^b\phi^b_{\ell}(\bm{\xi}^b_{q})\phi^b_j(\bm{\xi}^b_{q})\frac{\partial f_b^j}{\partial t}&=-\sum_{p,q}w_p^aw_q^b\frac{c_{ba}}{m_b}\frac{\partial\phi^b_{\ell}}{\partial\bm{u}'}\Big\vert_{\bm{\xi}^b_q}\cdot \mathbf{Q}(\bm{\xi}^b_q,\bm{\xi}^a_p)\cdot\bm{\Gamma}_{ba}(\bm{\xi}^b_q,\bm{\xi}^a_p),\quad \forall \ell \in I^b,
\end{align}
where the expression for the antisymmetric vector is
\begin{align}
    \bm{\Gamma}_{ab}(\bm{\xi}^a_p,\bm{\xi}^b_q)=\sum_{i,j}f_b^if_a^j\frac{\phi_i^b(\bm{\xi}^b_q)}{m_a}\frac{\partial \phi_j^a}{\partial\bm{u}}\Big\vert_{\bm{\xi}^a_p}-f_b^if_a^j\frac{\phi_j^a(\bm{\xi}^a_p)}{m_b}\frac{\partial \phi^b_i}{\partial\bm{u}'}\Big\vert_{\bm{\xi}^b_q}=-\bm{\Gamma}_{ba}(\bm{\xi}^b_q,\bm{\xi}^a_p).
\end{align}

To illustrate the necessary conditions for a conservative discretization, let us choose some numbers $\{\psi^a_k\}_{k\in I^a}$ and $\{\psi^b_{\ell}\}_{\ell\in I^b}$ which, for now, are arbitrary. We multiply the equations for the degrees of freedom with these numbers, sum them together, and concentrate on the resulting expression
\begin{multline}
    \sum_{p,i,k}w_p^a\psi^a_k\phi^a_k(\bm{\xi}^a_{p})\phi^a_i(\bm{\xi}^a_{p})\frac{\partial f_a^i}{\partial t}+\sum_{q,j,\ell}w_q^b\psi^b_{\ell}\phi^b_{\ell}(\bm{\xi}^b_{q})\phi^b_j(\bm{\xi}^b_{q   })\frac{\partial f_b^j}{\partial t}\\
    =\sum_{p,q}c_{ab}w_p^aw_q^b\left(\sum_{\ell}\frac{\psi^b_{\ell}}{m_b}\frac{\partial\phi^b_{\ell}}{\partial\bm{u}'}\Big\vert_{\bm{\xi}^b_q}-\sum_k\frac{\psi^a_k}{m_a}\frac{\partial\phi^a_k}{\partial\bm{u}}\Big\vert_{\bm{\xi}^a_p}
    \right)\cdot \mathbf{Q}(\bm{\xi}^a_p,\bm{\xi}^b_q)\cdot\bm{\Gamma}_{ab}(\bm{\xi}^a_p,\bm{\xi}^b_q),
\end{multline}
which follows from the antisymmetry of the vector $\bm{\Gamma}_{ab}$ and the symmetry of $c_{ab}$ and $\mathbf{Q}$. The left side represents the sum of collisional rates of change of the quantities $\sum_{k}\psi^a_k\phi^a_k(\bm{u})$ for species $a$ and $\sum_{\ell}\psi^b_{\ell}\phi^b_{\ell}(\bm{u})$ for species $b$. With any polynomial bases $\{\phi^a_i\}_{i\in I^a}$ and $\{\phi^b_i\}_{i\in I^b}$ one can represent global polynomial functions up to the same polynomial order as the bases. Hence the expressions $\sum_{k}\psi^a_k\phi^a_k(\bm{u})$ and $\sum_{\ell}\psi^b_{\ell}\phi^b_{\ell}(\bm{u})$ can exactly present the quantities $\psi_a=m_a\{1,u^x,u^y,u^z\}$ and $\psi_b=m_b\{1,u^x,u^y,u^z\}$ over the respective domains. For these specific functions one then finds that the expression
$$
\sum_{\ell}\frac{\psi^b_{\ell}}{m_b}\frac{\partial\phi^b_{\ell}}{\partial\bm{u}'}\Big\vert_{\bm{\xi}^b_q}-\sum_k\frac{\psi^a_k}{m_a}\frac{\partial\phi^a_k}{\partial\bm{u}}\Big\vert_{\bm{\xi}^a_p}
$$
vanishes exactly analogously to the infinite-dimensional case, demonstrating that a finite-element scheme with at least linear basis functions will automatically satisfy the number and momentum density conservation in both the nonrelativistic and relativistic case. Based on the above analysis, it would seem appropriate to expect the energy conservation to follow from similar steps: (i) find a way to represent the energies of species $a$ and $b$ with the coefficients $\{\psi^a_k\}_{k\in I^a}$ and $\{\psi^b_{\ell}\}_{{\ell}k\in I^b}$, and (ii) exploit the null space of the tensor $\mathbf{Q}$. In the nonrelativistic case, both conditions are achieved exactly in this manner, by using a second order polynomial basis. In the nonrelativistic case, both conditions, however, require additional work as the expression for energy is not a polynomial. It turns out that solving the two issues in the relativistic case also provides a recipe to achieve energy conservation in the nonrelativistic case using only linear elements.

{\it Key observation for energy conservation:}
We address the issue (ii) first, with the solution instructing us on how to approach the issue (i). Unraveling of the knot begins with the observation that the infinite-dimensional energy-conservation conditions in both the relativistic and nonrelativistic case, namely \eqref{eq:relativistic_null_space} and \eqref{eq:nonrelativistic_null_space}, can be expressed in terms of the respective gradient vectors of the particle energies. This follows from the seemingly meaningless rearrangements
\begin{align}
    \bm{u}&=\frac{\frac{\partial\gamma c^2}{\partial\bm{u}}}{\sqrt{1-\frac{1}{c^2}|\frac{\partial\gamma c^2}{\partial\bm{u}}|^2}}, && \text{relativistic},\\
    \bm{u}&=\frac{1}{2}\frac{\partial |\bm{u}|^2}{\partial \bm{u}}, && \text{nonrelativistic},
\end{align}
and then substituting these expressions into the corresponding tensors $\mathbf{Q}$, according to
\begin{align}
    \mathbf{Q}_{\text{BB}}(\bm{u},\bm{u}')&=\mathbf{Q}_{\text{BB}}\left(\frac{\frac{\partial\gamma c^2}{\partial\bm{u}}}{\sqrt{1-\frac{1}{c^2}|\frac{\partial\gamma c^2}{\partial\bm{u}}|^2}},\frac{\frac{\partial\gamma' c^2}{\partial\bm{u}'}}{\sqrt{1-\frac{1}{c^2}|\frac{\partial\gamma' c^2}{\partial\bm{u}'}|^2}}\right)\equiv\mathbf{Q}_{\text{BB}\gamma}\left(\frac{\partial \gamma c^2}{\partial \bm{u}},\frac{\partial \gamma'c^2}{\partial \bm{u}'}\right),\\
    \mathbf{Q}_{\text{L}}(\bm{u},\bm{u}')&=\mathbf{Q}_{\text{L}}\left(\frac{1}{2}\frac{\partial |\bm{u}|^2}{\partial \bm{u}},\frac{1}{2}\frac{\partial |\bm{u}'|^2}{\partial \bm{u}'}\right).
\end{align}
At this point, one realizes that the null spaces of the tensors $\mathbf{Q}_{\text{BB}\gamma}$ and $\mathbf{Q}_{\text{L}}$ can, in fact, be expressed in terms of arbitrary functions $h(\bm{u})$ and $g(\bm{u})$ to read
\begin{align}
&\left(\frac{\partial h}{\partial\bm{u}}-\frac{\partial g}{\partial\bm{u}'}\right)\cdot\mathbf{Q}_{\text{BB}\gamma}\left(\frac{\partial h}{\partial \bm{u}},\frac{\partial g}{\partial \bm{u}'}\right)=\bm{0},\\
&\left(\frac{\partial h}{\partial\bm{u}}-\frac{\partial g}{\partial\bm{u}'}\right)\cdot\mathbf{Q}_{\text{L}}\left(\frac{\partial h}{\partial \bm{u}},\frac{\partial g}{\partial\bm{u}'}\right)=\bm{0}.
\end{align}

The strategy to obtain energy conservation is then to approximate the particle energies with finite-element functions
\begin{align}
    \gamma \approx \sum_k\gamma^a_k\phi^a_k(\bm{u}), && \gamma'\approx \sum_{\ell}\gamma^b_{\ell}\phi^b_{\ell}(\bm{u}), && \text{relativistic} \\
    \frac{1}{2}|\bm{u}|^2\approx\sum_k\mathcal{E}^a_k\phi_k^a(\bm{u}), && \frac{1}{2}|\bm{u}'|^2\approx\sum_{\ell}\mathcal{E}^b_{\ell}\phi_{\ell}^a(\bm{u}), && \text{nonrelativistic}
\end{align}
and to make the following substitutions in the equations of motion \eqref{eq:equations_of_motion_fa} and \eqref{eq:equations_of_motion_fb}
\begin{align}
    \mathbf{Q}(\bm{\xi}^a_p,\bm{\xi}^b_q)\rightarrow
    \begin{cases}
    \mathbf{Q}_{\text{BB}\gamma}\left(\sum_i\gamma^a_ic^2\frac{\partial \phi^a_i(\bm{u})}{\partial \bm{u}}\Big\vert_{\bm{\xi}^a_p},\sum_{j}\gamma_j^bc^2\frac{\partial \phi_j^b(\bm{u}')}{\partial \bm{u}'}\Big\vert_{\bm{\xi}^b_q}\right), \qquad  \text{relativistic},
    \\
    \\
    \mathbf{Q}_{\text{L}}\left(\sum_i\mathcal{E}^a_i\frac{\partial \phi^a_i(\bm{u})}{\partial \bm{u}}\Big\vert_{\bm{\xi}^a_p},\sum_{j}\mathcal{E}_j^b\frac{\partial \phi_j^b(\bm{u}')}{\partial \bm{u}'}\Big\vert_{\bm{\xi}^b_q}\right), \qquad  \text{nonrelativistic}.
    \end{cases}
\end{align}
The expressions for the rate-of-change of energy density will then vanish identically as can be observed both in the relativistic case
\begin{align}
    &m_ac^2\sum_{p,i,k}w_p^a \gamma^a_k\phi^a_k(\bm{\xi}^a_{p})\phi^a_i(\bm{\xi}^a_{p})\frac{\partial f_a^i}{\partial t}+m_bc^2\sum_{q,j,\ell}w_q^b\gamma^b_{\ell}\phi^b_{\ell}(\bm{\xi}^b_{q})\phi^b_j(\bm{\xi}^b_{q   })\frac{\partial f_b^j}{\partial t} \nonumber\\
    &=\sum_{p,q}c_{ab}w_p^aw_q^b\left(\sum_{\ell}\gamma^b_{\ell}c^2\frac{\partial\phi^b_{\ell}}{\partial\bm{u}'}\Big\vert_{\bm{\xi}^b_q}-\sum_k\gamma^a_kc^2\frac{\partial\phi^a_k}{\partial\bm{u}}\Big\vert_{\bm{\xi}^a_p}
    \right)\nonumber
    \\
    &\qquad \cdot \mathbf{Q}_{\text{BB}\gamma}\left(\sum_i\gamma^a_ic^2\frac{\partial \phi^a_i}{\partial \bm{u}}\Big\vert_{\bm{\xi}^a_p},\sum_{j}\gamma_j^bc^2\frac{\partial \phi_j^b}{\partial \bm{u}'}\Big\vert_{\bm{\xi}^b_q}\right)\cdot\bm{\Gamma}_{ab}(\bm{\xi}^a_p,\bm{\xi}^b_q)=0,
\end{align}
and in the nonrelativistic case
\begin{align}
    &m_a\sum_{p,i,k}w_p^a \mathcal{E}^a_k\phi^a_k(\bm{\xi}^a_{p})\phi^a_i(\bm{\xi}^a_{p})\frac{\partial f_a^i}{\partial t}+m_b\sum_{q,j,\ell}w_q^b\mathcal{E}^b_{\ell}\phi^b_{\ell}(\bm{\xi}^b_{q})\phi^b_j(\bm{\xi}^b_{q   })\frac{\partial f_b^j}{\partial t}\nonumber\\
    &=\sum_{p,q}c_{ab}w_p^aw_q^b\left(\sum_{\ell}\mathcal{E}^b_{\ell}\frac{\partial\phi^b_{\ell}}{\partial\bm{u}'}\Big\vert_{\bm{\xi}^b_q}-\sum_k\mathcal{E}^a_k\frac{\partial\phi^a_k}{\partial\bm{u}}\Big\vert_{\bm{\xi}^a_p}
    \right)\nonumber\\
    &\qquad \cdot \mathbf{Q}_{\text{L}}\left(\sum_i\mathcal{E}^a_i\frac{\partial \phi^a_i(\bm{u})}{\partial \bm{u}}\Big\vert_{\bm{\xi}^a_p},\sum_{j}\mathcal{E}_j^b\frac{\partial \phi_j^b(\bm{u}')}{\partial \bm{u}'}\Big\vert_{\bm{\xi}^b_q}\right)\cdot\bm{\Gamma}_{ab}(\bm{\xi}^a_p,\bm{\xi}^b_q)=0.
\end{align}
These substitutions will not affect the conservation of number or momentum density, and lead to fully conservative schemes even with linear elements.

{\it Summary: }
It was previously thought that achieving an energy conserving finite-element scheme for the Beliav-Budker collision operator would be challenging. The thought was based on the fact that the relativistic kinetic energy of a particle cannot be expressed exactly with polynomial basis functions and that, at the time, the existing conservative finite-element discretization of the nonrelativistic operator relied on exact representation of the particle energy with a finite-element basis~\cite{Hirvijoki-Adams:2017PhPl}. After the reporting of an energy-conserving finite-difference scheme~\cite{shiroto-sentoku:2019arXiv}, it nevertheless became clear that a modification of the arguments of the tensor $\mathbf{Q}$ appearing in the collision operator would alleviate the previous difficulties. The modification we have introduced to the evaluation of $\mathbf{Q}$ is justified as it converges to the original expression in the limit that the finite-element mesh becomes infinitely dense. Even on a finite mesh our approximation is expected to be physically meaningful and accurate since the limiting behaviours of the relativistic energy in terms of the particle momentum are quadratic and linear at the small and large energies, respectively. 

The proposed solution of manufacturing a desired null space is expected to work also in discretizing the metriplectic formulation of the relativistic collision operator, akin to the works~\cite{Kraus-Hirvijoki-2017,Hirvijoki-Kraus-Burby:2018arXiv,Hirvijoki-Burby-Kraus:2018arXiv}. Verification of this is, however, left to future publications. As a final note, we would like to mention that the idea of manufacturing a null space is not new. A similar approach was used also in~\cite{Burby-Brizard:2015PhPl} and~\cite{Hirvijoki-Burby-2017} to derive a conservative collision operator for gyrokinetics and its metriplectic formulation. Why it took so long to realize the applicability of this trick to discretizations of the collision operator is a good question.

\bibliographystyle{apsrev4-1}
\bibliography{references}   

\end{document}